\documentclass[superscriptaddress,twocolumn,showpacs,preprintnumbers,amsmath,amssymb,prl]{revtex4}

\usepackage{graphicx}
\usepackage{dcolumn}
\usepackage{bm}

\begin{document}


\title{Successive phase transitions and phase diagrams of the quasi-two-dimensional triangular antiferromagnet Rb$_4$Mn(MoO$_4$)$_3$}

\author{Rieko Ishii}
\affiliation{Institute for Solid State Physics (ISSP), University of Tokyo, Kashiwa, Chiba 277-8581, Japan}
\author{Shu Tanaka}
\affiliation{Institute for Solid State Physics (ISSP), University of Tokyo, Kashiwa, Chiba 277-8581, Japan}
\author{Keisuke Onuma}
\affiliation{Department of Physics, Kyoto University, Kyoto 606-8502, Japan}
\author{Yusuke Nambu}
\affiliation{Department of Physics and Astronomy, Johns Hopkins University, Baltimore, MD 21218, USA}
\affiliation{NIST Center for Neutron Research, National Institute of Standards and Technology (NIST), Gaithersburg, MD 20899, USA}
\author{Masashi Tokunaga}
\affiliation{Institute for Solid State Physics (ISSP), University of Tokyo, Kashiwa, Chiba 277-8581, Japan}
\author{Toshiro Sakakibara}
\affiliation{Institute for Solid State Physics (ISSP), University of Tokyo, Kashiwa, Chiba 277-8581, Japan}
\author{Naoki Kawashima}
\affiliation{Institute for Solid State Physics (ISSP), University of Tokyo, Kashiwa, Chiba 277-8581, Japan}
\author{Yoshiteru Maeno}
\affiliation{Department of Physics, Kyoto University, Kyoto 606-8502, Japan}
\author{Collin Broholm}
\affiliation{Department of Physics and Astronomy, Johns Hopkins University, Baltimore, MD 21218, USA}
\affiliation{NIST Center for Neutron Research, National Institute of Standards and Technology (NIST), Gaithersburg, MD 20899, USA}
\author{Dixie P. Gautreaux}
\author{Julia Y. Chan}
\affiliation{Department of Chemistry, Louisiana State University, Baton Rouge, LA 70803, USA}
\author{Satoru Nakatsuji}
\affiliation{Institute for Solid State Physics (ISSP), University of Tokyo, Kashiwa, Chiba 277-8581, Japan}

\date{\today}

\begin{abstract}
Comprehensive experimental studies by magnetic, thermal and neutron measurements have clarified that Rb$_4$Mn(MoO$_4$)$_3$ is a model system of a quasi-2D triangular Heisenberg antiferromagnet with an easy-axis anisotropy, exhibiting successive transitions across an intermediate collinear phase. As a rare case for geometrically frustrated magnetism, quantitative agreement between experiment and theory is found for complete, anisotropic phase diagrams as well as magnetic properties. 
\end{abstract}

\maketitle{}
Geometrically frustrated magnetism has been a subject of active research in condensed matter physics. Generally in physics, quantitative comparison between experiment and theory is crucial to make a firm progress of our understanding. In the field of frustrated magnetism, however, such fortuitous cases are still scarce where full or semi quantitative agreement between experiment and theory have been found, except a few examples such as spin ice \cite{bramwell} and the orthogonal dimer SrCu$_2$(BO$_3$)$_2$ \cite{Kageyama3}. 

Two-dimensional (2D) triangular antiferromagnets (TAFM) have been extensively studied, because of rich frustrated magnetism expected on the simple 2D Bravais lattice \cite{Collins,Shimizu,Misguich,Nigas}. 
Theoretically, it has gained a consensus that the ground state for the nearest-neighbor antiferromagnetic (AF) Heisenberg model has the 120$^\circ$ spin order \cite{Huse,Bernu,Capriotti}. In this case, the concept of vector chirality, the handedness of the way the spins are rotated in a 120$^\circ$ order for a given triangle, may become essential and lead to exotic phenomena such as phase transitions with a new universality class \cite{KawamuraReview} and multiferroic phenomena~\cite{MultiferroicReview}.

In fact, the vector chirality is expected to play an important role in phase transitions on a triangular lattice, {\it{e.g.,}} for the nearest neighbor Heisenberg model:
\begin{eqnarray}
\mathcal {H}= J{\sum_{\langle i,j\rangle }}{\bm {S}_{i}} \cdot {\bm {S}_{j}}-D\sum_{i}{(S_{i}^{z})^2}-g\mu_{\rm B} \sum_{i}{\bm {H}\cdot \bm S_{i}},
\label{Hamiltonian}
\end{eqnarray}
where $J>0$ and $D$ are intralayer exchange interaction, and single-ion anisotropy, respectively. Depending on the anisotropy, namely, the sign of $D$, the model exhibits three types of phase transitions, featuring effects due to the vector chirality. For Heisenberg spin with $D=0$, the chirality forms a point defect called Z$_2$ vortex, and a proposal has been made on its binding-unbinding transition at finite temperature \cite{z2vortex}. 
For XY type ($D<0$), a long-range order of the vector chirality without dipole magnetic order is expected at slightly higher temperature than the Kosterlitz-Thouless transition into 120$^\circ$ quasi-long-range order \cite{MiyashitaShiba}. 
For easy axis case ($D>0$), successive transitions are expected associated with respective ordering of the longitudinal and transverse spin components \cite{Kawamura2,Miyashita}.  Namely on cooling, the system first forms a collinear intermediate phase (IMP) with the three-sublattice `uud' structure \cite{Zhitomirsky}, and then transits into a 120$^\circ$ spin-order phase with a uniform vector chirality. Recent theoretical study has found another phase transition in the IMP, which separates the lower $T$ `uud' phase and a higher $T$ collinear phase with three different sublattice moments. However, the latter phase is only stable in the purely 2D limit, and thus with finite interlayer coupling, the `uud' phase should become dominant throughout the collinear IMP \cite{Zhitomirsky}.

For this easy-axis case, unlike the Heisenberg and XY, experiments have confirmed the theoretical predictions. Namely, the successive transitions and/or 1/3 magnetization plateau have been observed for TAFMs with easy-axis anisotropy such as VCl$_2$ \cite{Kadowaki1}, $A$CrO$_2$ ($A$ = Li, Cu) \cite{Kadowaki2,Kimura}, and a metallic TAFM GdPd$_{2}$Al$_{3}$ \cite{Kitazawa}. However, neither detailed study of the phase diagram under external field nor quantitative comparison between experiment and theory has been made so far, because of a relatively large scale of $J$ and/or lattice deformation due to magnetostriction.  
\begin{figure*}[t]
\includegraphics[width=39pc]{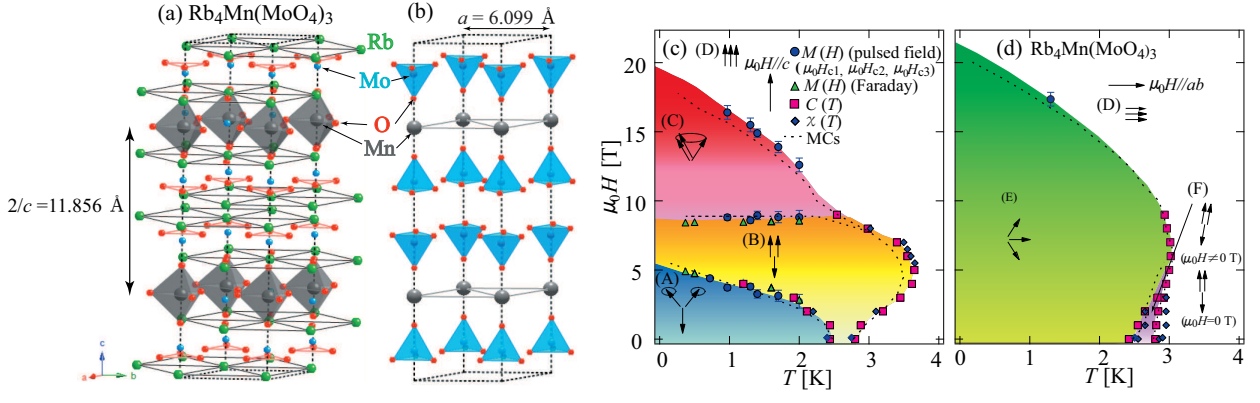}
\caption{(color online) Crystal structures of Rb$_4$Mn(MoO$_4$)$_3$ featuring (a) MnO$_5$ polyhedra, (b) equilateral triangular lattices of Mn$^{2+}$ and MoO$_4$ tetrahedra. Intralayer and interlayer distances between Mn$^{2+}$ ions are given by $a = 6.099$ \AA \ and $c/2 = 11.856$ \AA \, respectively, using the lattice constants at 298 K. Phase diagrams of Rb$_4$Mn(MoO$_4$)$_3$ for (c) $\mu_{0}H \parallel c$ and (d) $\mu_{0}H \parallel ab$ constructed by using various experimental techniques, and by Monte Carlo simulations (MCs) for $D/J$ = 0.22 (broken lines).\label{RbMn}}
\end{figure*}

Here, we report a comprehensive study on the crystal/spin structures and thermodynamic properties of the quasi-2D Heisenberg TAFM Rb$_4$Mn(MoO$_4$)$_3$ (RMMO). This material exhibits the successive transitions and $1/3$ magnetization plateau phase under field, reflecting its easy-axis anisotropy. The relatively small exchange coupling $J$ allows us to construct complete phase diagrams for the first time under field both parallel and perpendicular to the easy axis. As a rare case in geometrically frustrated magnets, quantitative agreement between experiment and theory has been found for the phase diagrams and magnetic properties, establishing the system as a model 2D Heisenberg TAFM characterized by the Heisenberg Hamiltonian of Eq. (1). 


Single crystals were synthesized by a flux method \cite{Solodovnikov}. The structure was determined by single crystal X-ray diffraction and adopts $P$6$_3$/{\it {mmc}} symmetry ($R_{1}$ = 2.88 \%).
Powder neutron diffraction (PND) measurements were performed on the BT1 at NIST, and confirmed that the structure is consistent with the X-ray results and stable down to 1.5 K.
Thus, RMMO has the equilateral triangular lattice formed by Mn$^{2+}$. Each Mn$^{2+}$ ion locates in a MnO$_5$ polyhedron and has a high spin $t_{\rm 2g}^{3}e_{\rm g}^{2}$ state, providing a $S=5/2$ Heisenberg spin. The dominant intralayer coupling $J$ should be made by the superexchange path Mn-O-O-Mn involving two oxygen atoms.
The interlayer interaction should be negligibly weak because of the large separation between Mn$^{2+}$ ion layers due to two Rb${^+}$ ions and two MoO$_4$ tetrahedra. 

D.c.\ $M$ was measured by a commercial SQUID magnetometer above 1.8 K, and by a Faraday method for 0.37 K $< T <2$ K \cite{sakakibara}. Specific heat $C_P$ was measured by a thermal relaxation method down to 0.4 K under fields up to 9 T. Pulsed field measurements of $M$ were performed up to 27 T. 
Classical Monte Carlo simulations (MCs) were made using the standard heat-bath method. 



As a guide to understand the results, we first present in Figs.\ \ref{RbMn}(c) and (d) $T$-$H$ phase diagrams consisting of six phases (A)-(F). Each symbol represents a transition point determined by different probes. In the following, we describe details of the experiments used to construct the phase diagrams, and compare them with theory. 

\begin{figure}[b]
\includegraphics[width=20pc]{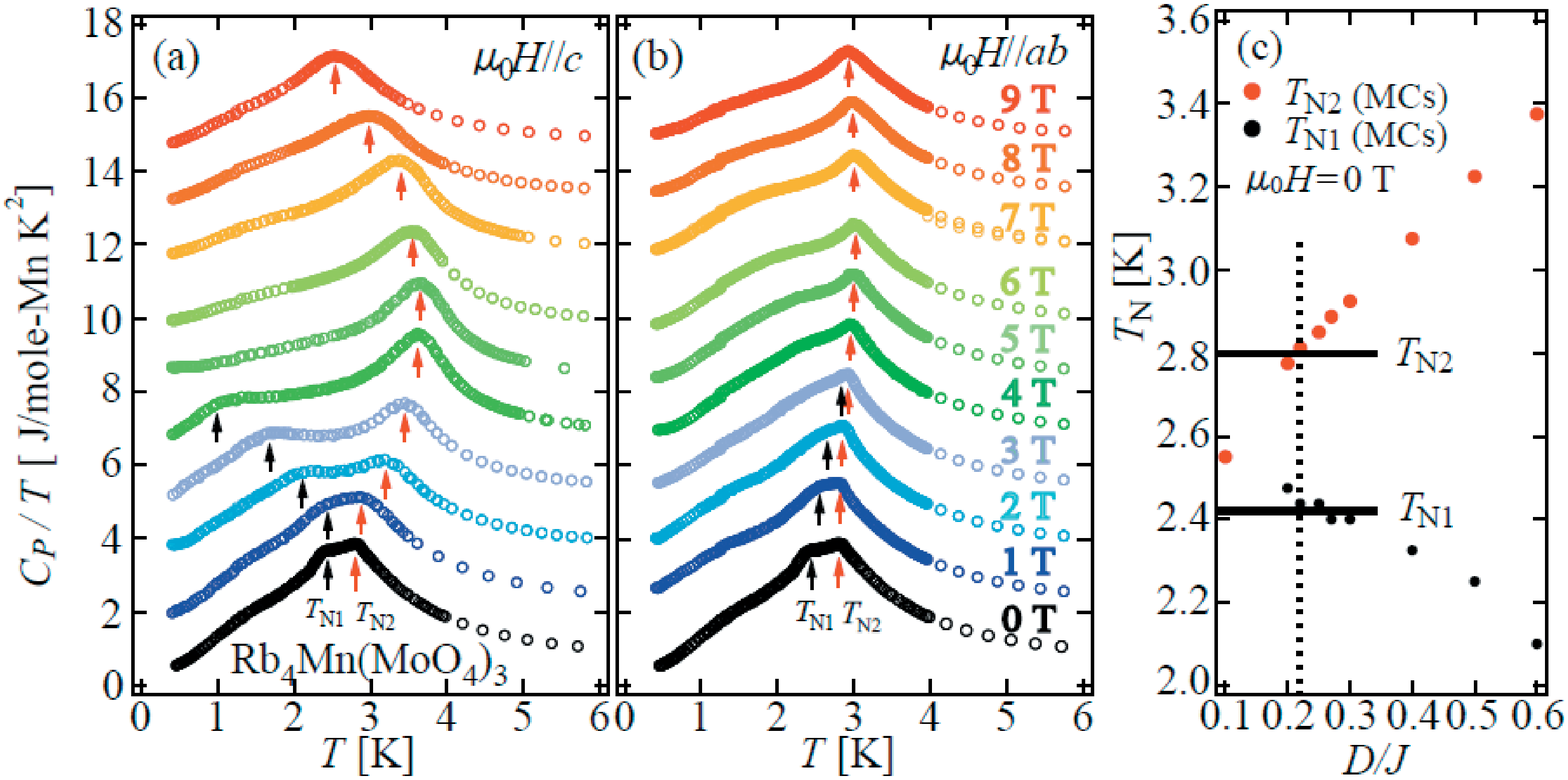}
\caption{(color online) $T$ dependence of $C_P/T$ under various fields for (a) $\mu_{0}H \parallel c$ and (b) $\mu_{0}H \parallel ab$. Values of $C_P/T$ under fields are shifted upwards for clarity. (c) $T_{\rm N1}$ (black arrow) and $T_{\rm N2}$ (red arrow) as a function of $D/J$ obtained by MCs for $J = 1.2$ K. The horizontal lines indicate $T_{\rm N1}$ = 2.42(2) K and $T_{\rm N2}$ = 2.80(2), obtained from $C_P/T$ at $\mu_0H = 0$. \label{SH_field}}
\end{figure}

We start with the specific heat measurement results. Notably, a double kink structure is found in the temperature dependence of the zero field $C_P$ (Fig.\ \ref{SH_field}). Furthermore, by applying field $\mu_{0}H \parallel c$, the two kinks become separated to form two peaks, providing clear evidence for the successive transitions. Here, we define $T_{\rm N1}$ and $T_{\rm N2}$ by the locations of the lower and higher temperature kinks or peaks, respectively. Under field $\mu_{0}H \parallel c$, the IMP in between $T_{\rm N1}$ and $T_{\rm N2}$ becomes stabilized, while under $\mu_{0}H \parallel ab$, $T_{\rm N1}$ approaches $T_{\rm N2}$ with increasing field, and finally the IMP disappears under $\mu_{0}H$ $>$ 4 T (Fig.\ \ref{SH_field}(b)). A broad tail of the peak seen at $T \ge T_{\rm N2}$ suggests an enhanced 2D spin fluctuations.

\begin{figure}[t]
\includegraphics[width=20pc]{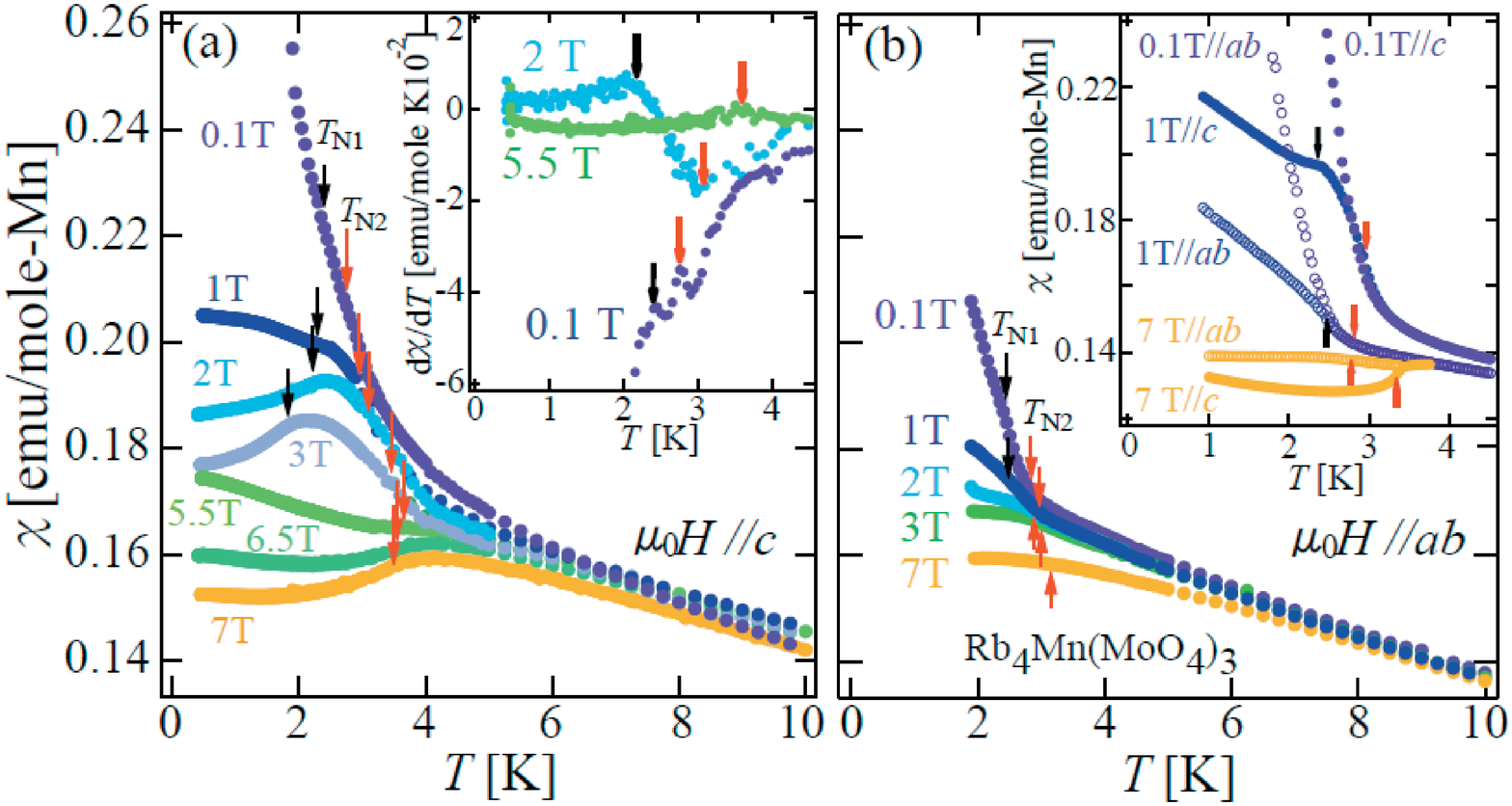}
\caption{(color online) (a) $c$-axis and (b) $ab$-plane components of the susceptibility $\chi(T)$ measured under various fields using zero-field-cooled sequence. Red and black arrows indicate $T_{\rm N1}$ and $T_{\rm N2}$ obtained by peaks in d$ \chi/$d$T$. Insets: (a) d$ \chi/$d$T$ vs. $T$ under $\mu_{0}H \parallel c$, (b) anisotropic $\chi(T)$ obtained by MCs for typical fields used in experiment shown in the main panels. \label{chi}}
\end{figure}

The successive phase transitions are also observed in the susceptibility $\chi \equiv  M/H$ (Fig.\ \ref{chi}), particularly as two peaks of the temperature dependence of ${\rm {d}}\chi/{\rm {d}}T$ under low fields (inset of Fig.\ \ref{chi}(a)). On the other hand, only a single peak is found in the high field regions of Phase (B) for $\mu_{0}H \parallel c$ and (E) for $\mu_{0}H \parallel ab$. These peak temperatures are indicated by arrows in Fig.\ \ref{chi} and are found to overlap the phase boundary given by the specific heat anomalies, as shown in Figs.\ \ref{RbMn}(c) and (d). The temperature dependence of both $\chi$ and $C_P/T$, and the slope of the phase boundary are found consistent with Ehrenfest's relation defined by $\Delta ((\partial {M}/\partial {T})_{H}/{\rm d}{T}) = -{\rm d}{T_{\rm N}}/{\rm d}H\Delta(C_P/T)$.  

The high temperature susceptibility shows the Currie-Weiss behavior. The fitting using the formula $C/(T + \Theta_{\rm W})$ in the range of 50-350 K yields the effective moment $p_{\rm eff}$ = 5.95 (5.97) $\mu_{\rm B}$ for $\mu_{0}H \parallel ab (c)$, close to the expected value 5.92 $\mu_{\rm B}$ for $S$ = 5/2, and AF Weiss temperature $\Theta_{\rm W}$ = $-$20.3(5) ($-$19.6(5)) K for $\mu_{0}H \parallel ab (c)$. The average $\Theta_{\rm W} \approx -20$ K corresponds to $J$ = 1.14(5) K.
The frustration parameter $|\Theta_{\rm W} |$/$T_{\rm {N2}}$ = 7.14 is large and indicates geometrical frustration. The anisotropic ratio $\chi_{c}/\chi_{ab}$, which is constant and near unity at high temperatures, becomes larger than unity below $\sim 10$ K, suggesting that RMMO has the easy-axis anisotropy.

\begin{figure}[t]
\includegraphics[width=19pc]{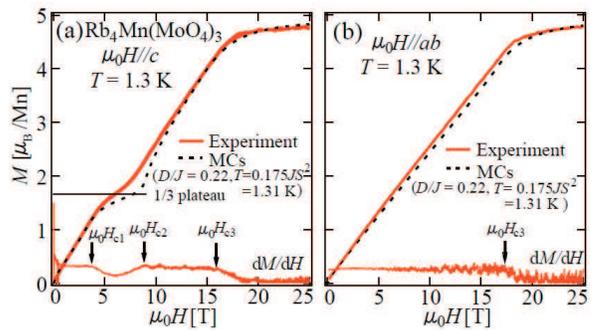}
\caption{(color online) Field dependence of the magnetization and its derivative for Rb$_4$Mn(MoO$_4$)$_3$ at $T$ = 1.3 K measured under a pulsed field (solid lines) and calculated using MCs (broken lines) for (a) $\mu_{0}H \parallel c$ and (b) $\mu_{0}H \parallel ab$. 
\label{MH}}
\end{figure}

Clear evidence for the easy-axis anisotropy is found in the field dependence of $M$. Figures \ref{MH} presents the $M(H)$ curve measured at 1.3 K as well as the results obtained by MCs. At $T$ $<$ $T_{\rm {N2}}$, a magnetic plateau was observed under $\mu_{0}H \parallel c$ at $\sim 1/3$ of the saturation moment of 5 $\mu_{\rm B}$ for Mn$^{2+}$ (Fig. \ref{MH}(a)). Meanwhile, for $\mu_{0}H\parallel ab$, $M$ increases monotonically with field and fully saturates to $\sim 5 \mu_{\rm B}$ (Fig. \ref{MH}(b)). Quantitative agreement with the MCs results is found except a slight deviation around the plateau region for $\mu_{0}H \parallel c$. Because $T \sim J$, the finite $T$ effect broadens the anomalies at the critical fields. Therefore, by using the kink found in d$M$/d$H$, we define the lower and upper critical fields of the plateau region, $\mu_{0}H_{\rm c1}$ and $\mu_{0}H_{\rm c2}$, and the critical field associated with the saturation of magnetic moment, $\mu_{0}H_{\rm c3}$. The 1/3 plateau of the $c$-axis $M(H)$ as well as its larger slope than the $ab$-plane $M(H)$ clearly indicates the easy-axis anisotropy. 
As can be seen in the phase diagram of Fig.\ \ref{RbMn}(c), the plateau field region found by the $M(H)$ curve becomes systematically wider with increasing temperature, and is smoothly connected with the IMP found by the temperature dependence of both $C_P/T$ and $\chi$. This provides experimental evidence that Phase (B) has the `uud' structure (Fig.\ \ref{RbMn}(c)).

By comparing experiment with theory, we first estimate the parameters $J$ and $D$. The mean-field theory predicts that the $ab$-plane and $c$-axis $M$ are linear in $H$ and follow $M = N_{\rm A}(g\mu_{\rm B})^2SH/(9J-2D)$ up to $\mu_{0}H_{\rm c3}$ and $M = N_{\rm A}(g\mu_{\rm B})^2SH/(9J-6D)$ up to $\mu_{0}H_{\rm c1}$, respectively \cite{Miyashita}. The experimental results in Figs.\ \ref{MH}(a) and (b) are indeed $H$-linear and the fits give $J$ = 1.2 K, $D$ = 0.28 K and thus $D/J = 0.23$. $J$ has nearly the same value as the one obtained from $\Theta_{\rm W}$. Then fixing $J = 1.2$ K, and comparing zero field $T_{\rm N1}$ and $T_{\rm N2}$ with those obtained by MCs as a function of $D/J$, we estimate $D/J$ to be 0.22(2), as indicated by a vertical broken line of Fig.\ \ref{SH_field}(c), consistent with the above estimate. 
Thus throughout the paper, we adopt $J = 1.2$ K and $D/J$ = 0.22.

Now, we compare the phase diagrams obtained by experiment and by the MCs performed for the 2D Heisenberg TAFM model represented by Eq. (1) (Figs.\ \ref{RbMn}(c) and (d)). The agreement between experiment (symbols) and theory (broken lines) is significantly good in detail for both field directions including the magnetic plateau phase. This suggests that the spin Hamiltonian for RMMO is well captured by Eq. (1), and the possible extra terms, for example, Dzyaloshinsky-Moriya interaction, should be negligibly small in comparison with the $J$ and $D$ terms. Three-sublattice spin structures inferred from both experiment and theory are schematically presented by solid arrows in each region of Phases (A)-(F).

At the ground state, the theory predicts a 120$^\circ$ structure with slight canting toward the $c$-axis due to the easy-axis anisotropy, as schematically shown in Phase (A) of Fig.\ \ref{RbMn}(c). The canting angle $\theta$ is estimated 2.7$^\circ$ for $H=0$ by the relation $\cos (\pi/3 - \theta) = 3J/(6J-2D)$ \cite{Miyashita}. This canting causes an increase in $\chi(T)$ below $T_{\rm N1}$, as consistently seen in both experiment and theory (Figs.\ \ref{chi}(a) and \ref{chi}(b) inset). According to theory, this canted structure is stable under $\mu_{0}H \parallel c$ in Phase (A). With further increasing $\mu_{0}H \parallel c$, however, the `uud' structure with a 1/3 magnetization plateau takes over in Phase (B) as observed in experiment, and then should transit into the `oblique' phase (C) with two parallel spins and one pointing to a different direction. Finally in Phase (D), the moments become fully polarized. 

Because $D/J \ll 1$, the ground state canted structure at $H=0$ should be nearly degenerate with a fan-shape structure having one spin in the $ab$-plane \cite{Miyashita}. This fan-shape structure has a weak ferromagnetic component in the $ab$-plane, and can be easily stabilized under a weak in-plane field in Phase (E), inducing an enhancement of $\chi(T)$ below $T_{\rm N1}$. For this behavior as well, we find a quantitative agreement between experiment and theory (Fig.\ \ref{chi}(b) and its inset). 
Notably, the fan-structure in Phase (E) a uniform vector chirality pointing to an in-plane direction perpendicular to the field. 
Under $\mu_{0}H \parallel ab$, the IMP(F) becomes narrower in temperature where the theory predicts that a collinear spin structure becomes inclined toward the field direction (Fig.\ \ref{RbMn}(d)). 

\begin{figure}[t]
\includegraphics[width=20pc]{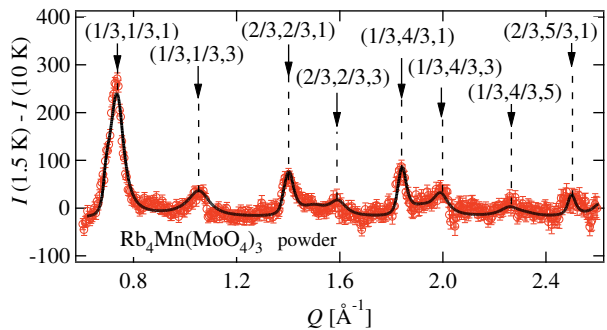}
\caption{(color online). Difference between the powder neutron diffraction obtained at 1.5 K and 10 K (red) together with the calculation for the quasi-2D 120$^\circ$ spin structure (black). \label{ND}}
\end{figure}
In order to confirm the 120$^\circ$ structure under zero field, we performed the PND measurements on the BT7 at NIST. Figure \ref{ND} shows the difference between the PND spectra obtained at 1.5 K and 10 K. A magnetic peak at the wave vector $\sim$ (1/3, 1/3, 1) indicates the 120$^\circ$ in-plane spin structure as well as AF interlayer correlations. 
We fit the data to the analytical formula for the spherical average of magnetic scattering from a quasi-2D 120$^\circ$ magnetic structure with tiny canting component corresponding to Phase (A). The out-of-plane correlations are described by Lorentzian and the in-plane correlations by Lorentzian squared \cite{Nigas}.
Consequently, the in-plane and out-of-plane correlation lengths are estimated to be $\xi_{ab}\gtrsim 98(1)$ \AA\ $\approx 16a$ \ and $\xi_{c} \approx 16.3(5)$ \AA\ $\approx 0.7c$, respectively, indicating quasi-2D magnetism and weak interlayer correlations. While further neutron experiments using a single crystal are necessary to more precisely determine the correlation function forms and lengths, the present observation provides a basis for the comparison between experiment and theory.

Our study has established a rare case of quantitative agreement between experiment and theory for geometrically frustrated magnetism. Our material should serve as a model system to further study novel 2D frustrated magnetism, such as critical dynamics associated with vector chirality, multiferroic noncolliner magnetism, and the possible magnon decay in the dispersion spectrum \cite{Zhitomirsky2}. 

This work is partially supported by Grant-in-Aid for Scientific Research (No.\ 19052004,\ 19340109,\ 21684019,\ 21840021) from JSPS, by Grant-in-Aid for Scientific Research on Priority Areas (No.\ 17071003,\ 19052003,\ 17072001) from MEXT, Japan, and by US-Japan Cooperative Program, ISSP. The computation is executed on computers at the Supercomputer Center, ISSP.

\end{document}